 \documentclass[aps,prb, preprint, superscriptaddress]{revtex4}

\usepackage{graphicx}
\newcommand\dt{\mbox{$\tilde{d}$}}
\newcommand\fp{\mbox{$f_{\rm pk}$}}
\def\fig#1{Figure \ref{fig#1}}

\begin{document}



\title{ Unequal Layer Densities in Bilayer Wigner Crystal at High Magnetic Field}



\author{Zhihai Wang}
\altaffiliation[Current address: ]{Riken, Advanced Science Institute, 2-1, Hirosawa,Saitama 351-0198, Japan }
\affiliation{National High Magnetic Field Laboratory, 1800 E. Paul Dirac Drive, Tallahassee, FL 32310}
\affiliation{Department of Physics, Princeton University, Princeton, NJ 08544}
\author{Yong P. Chen}
\altaffiliation[Current address: ]{Purdue University, West Lafayette, IN 47907}
\affiliation{National High Magnetic Field Laboratory, 1800 E. Paul Dirac Drive, Tallahassee, FL 32310}
\affiliation{Department of Electrical Engineering, Princeton University, Princeton, NJ 08544}
\author{Han Zhu}
\affiliation{National High Magnetic Field Laboratory, 1800 E. Paul Dirac Drive, Tallahassee, FL 32310}
\affiliation{Department of Physics, Princeton University, Princeton, NJ 08544}
\author{L. W. Engel}
\affiliation{National High Magnetic Field Laboratory, 1800 E. Paul Dirac Drive, Tallahassee, FL 32310}
\author{D. C. Tsui}
\affiliation{Department of Electrical Engineering, Princeton University, Princeton, NJ 08544}
\author{E. Tutuc}
\altaffiliation[Current address: ]{The University of Texas, Austin, TX 78712}
\affiliation{Department of Physics, Princeton University, Princeton, NJ 08544}
\author{M. Shayegan}
\affiliation{Department of Electrical Engineering, Princeton University, Princeton, NJ 08544}


\date{\today}

\begin{abstract}
  We report studies of pinning mode resonances of magnetic field induced bilayer Wigner crystals   of bilayer hole samples with negligible interlayer tunneling and different interlayer separations $d$,  in states with varying layer densities, including  unequal layer densities.  With unequal  layer densities, 
 samples with large $d$ relative to the in-plane carrier-carrier spacing $a$,   two pinning resonances   are present,  one for each layer.  For  small $d/a$ samples,  a single resonance is observed  even with significant  density imbalance.  These samples, at balance, were shown to exhibit an enhanced pinning mode frequency  [Zhihai Wang {\em et al.}, Phys. Rev. Lett. 136804 (2007)], which was  ascribed to a  one-component, pseudospin ferromagnetic Wigner solid.    The evolution of the resonance  frequency and line width indicates the quantum interlayer coherence survives at moderate density imbalance, but disappears when imbalance is sufficiently large.
\end{abstract}


\maketitle



\section{Introduction}

In high magnetic field $(B)$    two parallel layers of two-dimensional electrons in close proximity  have  been shown to exhibit many-body states with distinct quantum properties that  arise from  the   degree of freedom that identifies the layer.   
Explicit interlayer correlations are present in the   fractional quantum Hall state at total Landau level filling $\nu=1/2$  \citep{Suen-1992-PRL,Eisenstein-1992-PRL, Halperin-1983-ACTA, Yoshioka-1989-PRB, He-1991-PRB} as well as the celebrated  bilayer excitonic condensate  state at   $\nu=1$ \citep{Murphy-1994-PRL,lay94,Spielman-2000-PRL, Spielman-2001-PRL, Spielman-2004-PRB,eisnu1,tutucdrag,Yoshioka-1989-PRB, Fertig-1989-PRB, MacDonald-1990-PRL, He-1991-PRB}. The $\nu=1$ excitonic condensate state is seen    for   small   layer separation and an interlayer  sufficiently large that  interlayer tunneling is negligible.  
In the well-known language in which layer index is taken as  pseudospin, the state is  an   easy-plane   ferromagnet, and   has carrier wave functions  coherently distributed between the two layers.   Imbalance  \cite{tutucimbal,wiersma}, or unequal density for the two layers has    been shown \cite{tutucimbal} to strengthen the excitonic condensate state for moderate imbalance. 
%
%
 
Bilayers, like single layer 2D carrier systems, become insulating  at low enough $\nu$, and in   bilayers  as in  single layers\cite{msreview}, the  low $\nu$ insulators are understood  as  pinned Wigner solids \citep{suen92b,Suen-1994-SURF, Manoharan-1996-PRL, tutucimbal,faniel}.    Several  different bilayer Wigner crystal (BWC) states have been predicted theoretically  \citep{Esfarjani-1995-JPHYS, zhengfertig,ho,peeters} for the balanced condition, in which the carrier densities of the two layers are equal, and without interlayer tunneling (the relevant case for samples considered in this paper).  These phases depend on the relative strength of interlayer and intralayer carrier-carrier interactions, as determined by $d/a$, where $d$ is the interlayer separation, and    $a=( \pi p_{TOT})^{-1/2}$ is   the mean in-plane carrier spacing for total density $p_{TOT}$.    For the smallest $d/a$, a one-component phase  is predicted, which at each site of a triangular lattice  has a   carrier  coherently  present in both layers.  The one-component lattice   is an easy-plane pseudospin ferromagnetic BWC (FMBWC).  At larger $d/a$, without interlayer tunneling, the phases are two-component, with  each layer having  a lattice. The lattices are staggered with respect to each other, so the state is a   pseudospin antiferromagnetic BWC (AFMBWC).   Among these two-component staggered phases, a  square lattice was predicted to cover a wide  range of $d/a$ \citep{zhengfertig,ho,peeters}, though other phases, with rectangular or rhombic lattices, were also predicted \citep{ho,peeters}.   For large $d/a$, the intralayer interaction dominates, and each layer   has a triangular lattice like that of a single layer Wigner crystal.

  In both single layer \cite{murthyrvw,yewc,yongab,clidensity} and bilayer \cite{zhw,doveston}  systems, the spectra of the  pinned low $\nu$ solid exhibit striking resonances  that are understood as pinning modes,  in which pieces of the solid oscillate within the potential of  the residual disorder.  Useful information on the pinned solid is provided by the resonance:  its frequency  characterizes the strength of the pinning, and its integrated amplitude  indicates the number of participating carriers.    

This paper expands on our  earlier work  \cite{zhw}, which exploited the pinning  mode to characterize low $\nu$ BWC phases in samples at balance (with equal layer densities), using a series of  p-type samples  with essentially zero tunneling between layers, and  different layer separations. As in the earlier paper, we use   the parameter $\tilde{d}=2^{1/2} d/a$ to characterize effective separation.   The factor of $2^{1/2}$ is included in \dt\ for  comparison with the results \cite{Murphy-1994-PRL, Spielman-2000-PRL, Spielman-2001-PRL, Spielman-2004-PRB,eisnu1,tutucdrag} on the $\nu=1$ excitonic Bose condensate,  $\tilde{d}=d/l_{B1}$ where $l_{B1}$ is  the magnetic length at total Landau filling $\nu=1$ for a balanced state.    
Our  earlier paper\cite{zhw} described the  change in the pinning mode on changing the top and bottom layer densities,    from a balanced state  $(p,p)$ to a one-layer state $(p,0)$, in order to capture the effects of interlayer interaction and correlation on the balanced state.  
We found, for low interlayer separation, $\dt\le 1.8$, that the pinning mode resonance peak frequency, \fp\ was enhanced above the value expected \cite{clidensity} from the overall density, which determines the stiffness of the solid.   We interpreted the results as consistent with these low \dt\ samples being in an FMBWC phase at balance.  The interpretation is based on   theory \cite{yongtheory} which explains  an enhancement of \fp\    specific to the  FMBWC,  in the presence of disorder that is correlated in the  two layers.

In this paper we study the same series of samples with various \dt, varying the layer densities to look at unbalanced states.
We find that  on small imbalance, the resonance does not split for  samples with $\dt\le 1.8$.      For larger 
\dt, the resonance does split into two peaks,  one of which is mainly sensitive to the top   layer density, the other to the bottom layer density.    Moreover, the \fp\ enhancement observed earlier in the $\dt\le 1.8$ at balance persists out to significant imbalance.   We interpret these observations in the framework of FMBWC for low \dt\ samples, and AFMBWC for higher \dt\ samples. 

\begin{table}[htdp]
\begin{center}
\begin{tabular}{|c|c|c|c|c|c|}
\hline
Wafer & $d$ (\AA) & $p$ &  $\tilde{d}$ & $\nu=1$ & $p_B^{\ast}$ \\
Name &  & $(10^{10}{\rm cm}^{-2})$ &  & QHE & $(10^{10}{\ rm cm}^{-2})$ \\
\hline
$M440$ & 225 & 3.0 & 1.39 & yes & $>7$ \\
\hline
$M440$ & 225 & 3.85 & 1.57 & yes & $>7$ \\
\hline
$M465$ & 230 & 3.65 & 1.56 & yes & $>7$ \\
\hline
$M417$ & 260 & 3.05 & 1.61 & yes & $\sim4.8$ \\
\hline
$M433$ & 300 & 2.52 & 1.7 & no  & $\sim4.8$ \\
\hline
$M433$ & 300 & 2.85 & 1.8 & no  & $\sim4.7$ \\
\hline
$M436$ & 450 & 2.4 & 2.5 & no  & $ \sim1.7$ \\
\hline
$M443$ & 650 & 2.85 & 3.9 & no  & $\approx0$ \\
\hline
$M453$ & 2170 & 5.25 & 18 & no & $\approx0$ \\
\hline
\end{tabular}
\end{center}
\caption{All the samples discussed in this paper. For each sample, interlayer separation ($d$), per layer density at balance ($p$), and $\tilde{d}\equiv{d}/(4\pi{p})^{-1/2}$ are labeled. The samples whose balanced states show a $\nu=1$\  interlayer phase coherent quantum Hall state are marked.   For bottom layer density $p_B<p_B^{\ast}$, the BWC has a single pinning mode. $p_B^{\ast}$ is given for each sample, while for   $p_B>p_B^{\ast}$, the BWC has two pinning modes (except when the bilayer is near balance, see text).   The two lines each in the table for M440 and for M433 
 correspond to  different cool downs.}
\label{default}
\end{table}

 		\section{Experimental method}

We present results from p-type  GaAs/AlGaAs/GaAs    double quantum wells  (DQW) grown on (311)A substrates.   Designed  to suppress interlayer tunneling,  pieces  of the same wafers were studied   in earlier dc measurements focusing on $\nu=1$ \cite{tutucdrag,tutucimbal,tutucdiss}.     Table 1 summarizes the DQW dimensions and densities.    $d=w+b$ is the distance between the centers of the two QW's, where $b$ is the barrier width, and $w=150$\AA~ is the width of  each QW in all these samples.   $p$ is  the as-cooled top-layer density,  taken for zero  gate voltage.  
   $p_B^{\ast}$, to be discussed later, is the  observed   bottom-layer density, above which two pinning modes can be resolved in imbalanced states. M465 and M453 are asymmetrically doped on both sides of the DQW; all  other samples have dopants only at the front of the DQW.    The two lines for  M440    correspond to two different cool downs, each with different $p$. The interlayer barrier layer is AlAs for $d\le300$ \AA, but is a combination of AlAs and AlGaAs for $d>300$ \AA.


%

Our microwave measuring  setup has been described in  earlier publications \cite{murthyrvw,clibdep,yewc,yongab,clidensity}.    \fig{schm} shows a schematic of a bilayer sample and the measuring circuit.   A  metal film transmission line on  the top surface of the sample  couples capacitively with the bilayer.   The transmission line is  of the coplanar waveguide 
type, with narrow, driven center conductor, separated from broad grounded side planes by slots of width $W$. 
$|\sigma_{xx}|$ is small enough, and the operating frequency, $f$, is high enough, that the response of the bilayer 
  has  small   effect on the microwave electric field, so that this field is essentially the same in both layers.   In this 
  high-frequency, low-loss case, the in-plane microwave electric field is  well-confined to the region of bilayer immediately under the slots, and  
 we calculate   Re$(\sigma_{xx})=-W|\ln(P/P_0)|/2Z_0L$, where $P$ is the transmitted power, $P_0$ is the transmitted power for a reference state  in which bilayer conductivity is zero, $Z_0=50\ \Omega$ is the characteristic impedance calculated from the transmission line geometry for $\sigma_{xx}=0$, and $L$ is the length of the line. 
The data  were obtained in the limit of low applied power; we verified that reducing power   further did not affect the measured Re$(\sigma_{xx})$.   

The samples have ohmic  contacts on the edges, which connect the two layers together and allow 
gate biasing of the bilayer. The microwave measurement precludes a conventional front gate for independent control of top and bottom layer densities; such a gate in the slots of the  
coplanar waveguide would effectively short circuit the  transmission line.   Hence, unlike in bilayer dc transport studies 
such as in refs. \cite{Suen-1992-PRL, Suen-1994-SURF, Eisenstein-1992-PRL, lay94,Manoharan-1996-PRL,tutucdrag,tutucimbal,Murphy-1994-PRL, Spielman-2000-PRL, Spielman-2001-PRL, Spielman-2004-PRB}, we control the layer densities only by back gate bias relative to the bilayer. 
One  consequence of this is that the only  accessible balanced   state is    $(p,p)$, and it is for this state that \dt\ is calculated.

 Determination of the top and bottom layer densities $p_T$ and $p_B$  is based on   the position and behavior of IQHE minima in  Re$(\sigma_{xx})$ vs $B$ curves taken at 200  MHz. 
 For the smaller $d$ samples (M440, M465, and M417), the integer quantum Hall effect (IQHE), including the $\nu=1$ QHE,  is observed for  several integer total filling factors $\nu$, allowing  the total density $p_{TOT}$ vs $V_g$ to be obtained. The balanced condition  can be found easily from the  development of the  $\nu=1$ and $\nu=2$ IQHE's  , which 
are respectively weakest or strongest in the  balanced state $(p,p)$.      $p_T$ and $p_B$  vs  back gate voltage $V_g$ are found using a simple capacitive model, whose main parameter, $C_g$, is the capacitance per unit area which  relates   change in $p_{TOT}$   to a change in $V_g$ through   $\Delta p_{TOT}=C_g \Delta V_g$.   In this model, both  2D layers    are neglected when  fully depleted but are considered as ideal conductors (large density of states) otherwise.   The model has two cases: 1)  $p_B>0$,   $\Delta p_B=C_g \Delta V_g$  while $p_T$ is constant and equal to its as-cooled, unbiased  value,  $p$,  and 2)  $p_B=0$ and $\Delta p_T=C_g  \Delta V_g$. 
 For the larger $d$ samples (M433, M436, M443, and M453), the top and bottom layer densities can be derived from the positions of the IQHE minima associated with individual layers, but only for $V_g$ ranges in which such minima are resolved. For $V_g $ outside of  these ranges, the densities are estimated by extrapolation, also based on the capacitive model.

The capacitive model neglects kinetic energy and carrier interaction in the layers.   For a hole bilayer with a backgate,  when $p_B$ is nearly depleted, this can cause $p_T$ to increase slightly with increasing $V_g$ (or decreasing $p_{TOT}$), while $p_B$ decreases  faster than linearly.
This deviation from the capacitive model, referred to as carrier transfer, was studied $B=0$ \cite{Ying-1995-PRB, Papadakis-1997-PRB, Eisenstein-1994-PRB}, and considered in the high $B$ insulating regime \cite{Eisenstein-1994-PRB}.  
Carrier transfer does not affect the relationship of $p_{TOT}$ and $V_g$, and does not affect the identification of the  balanced state $(p,p)$.   Following  calculations in refs. \cite{Ruden-1991-APL, Katayama-1995-PRB},
using a classical estimate of a single-layer Wigner crystal cohesive energy, and neglecting the kinetic energy since both layers are considered in the lowest Landau level,  we find the capacitive model can  underestimate the the $p_{TOT}$ at which the the bottom layer is completely depleted  by as much as $1\times10^{10}$ cm$^{-2}$.   

\begin{figure}
\centering
\includegraphics[width=7.5cm]{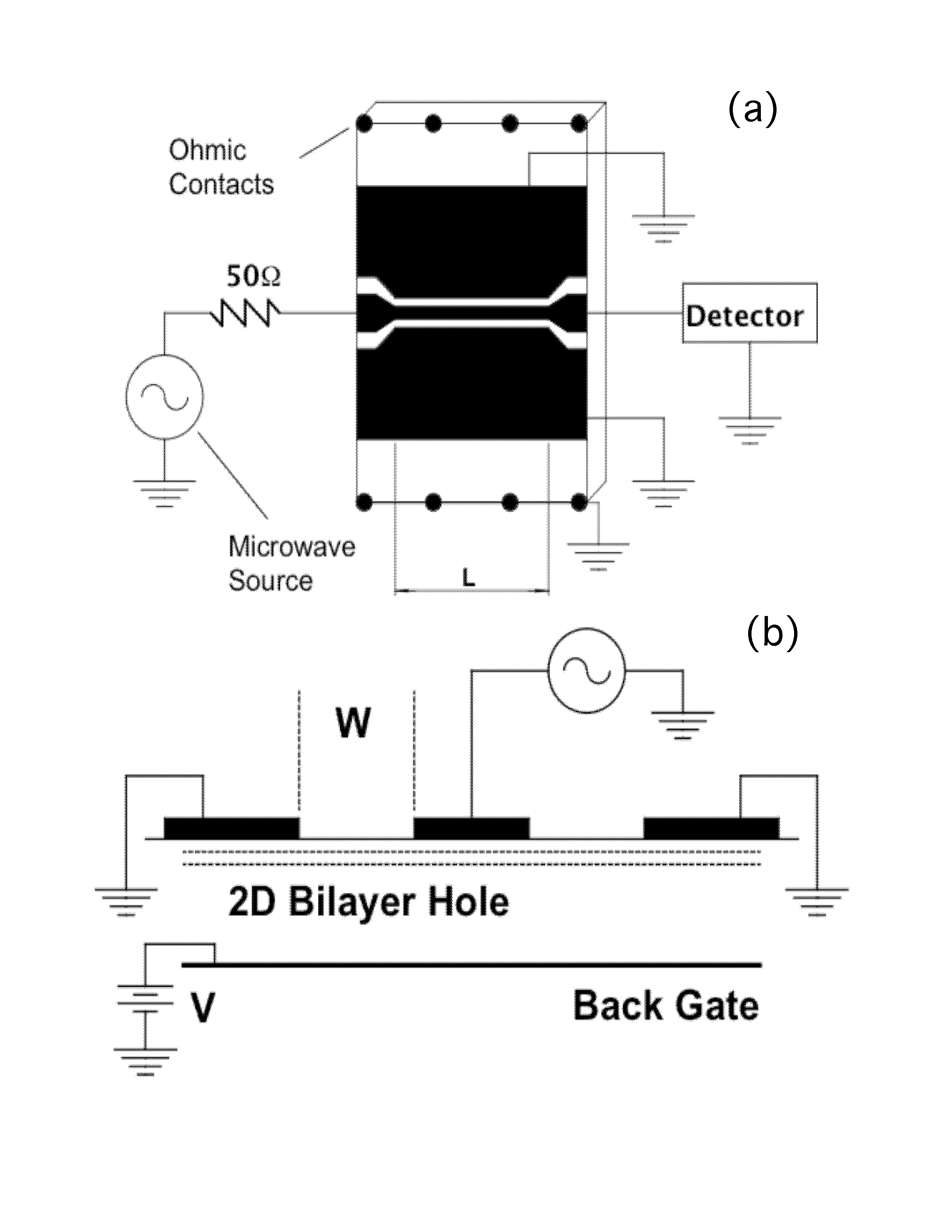}
\caption{Schematic representation of microwave circuit used in our measurement, not to scale. (a) top-view. Black areas represent metal films on the sample surface. (b) Side-view. Heavy black lines indicate metal film of the transmission line structure.  } 
\label{figschm}
\end{figure}

%
 
\section{Experimental Results}
		 

%

 
 \subsection{Large separation}
 For the most widely separated samples,  M453 and M443, we find two resolved pinning modes whenever there is significant $p_B$, as long as the layer densities  are not too close to balance.   M453 represents the independent layer case, with   $d=2170$\AA~ and $\tilde{d}=18$.     \fig{spx453}a shows spectra as $B$  increases   for  a state with layer densities 
 $(p_T,p_B)=(5.3,3.3)\times 10^{10}$ cm$^{-2}$.  At high $B$, two distinct peaks are visible in the spectrum, which are labeled   ``T" and ``B" on the graph, since they are identified with the top and bottom layers of the sample. 
  The peaks appear  as the filling factor of the respective layer goes below about 0.3, consistent with an earlier 
  study \cite{clibdep} of  pinning mode development  in single-layer p-type samples.     In the rest of this paper 
  we are concerned only with this low $\nu$ solid regime, in which the pinning modes  are well developed.    
  
     \fig{spx453}b  shows  spectra of M453  at $14$ T   in bilayer states with fixed $p_T$ and varying $p_B$, as required by the experimental constraints   discussed in the previous section.  The behavior is consistent with the independent layer case and with the association of each resonances with a particular layer.  Resonance T remains at fixed frequency and intensity as  $p_B$ is varied, while   resonance B increases in peak frequency, $f_{pk}$ as $p_B$ is decreased.  The B resonance  vanishes as $p_B$ approaches zero, and the two resonances overlap for the balanced state.  

  M443 has $d=650$ \AA~ and $\tilde{d}= 3.9$, so its layers are more closely coupled than those of M453.   
\fig{spx443} shows a set of spectra with fixed $p_T$ and varying $p_B$, for sample M443 at $B=9$ T.   Again one resonance, which we label ``B", is much more sensitive to $p_B$, but in this case the other resonance, which we label ``T",  
has a slight but measurable $p_B$ dependence, demonstrating that there is at least some dynamical coupling between the layers.     For $p_B<p_T$, resonance  T  dominates and  resonance  B  shows as a high $f$ shoulder. 
 
\fig{sepsum}  summarizes the data for  M453 and M443,  which have well-resolved pinning modes   for all significant $p_B$, as long as the  bilayer is away from balance.  The dependence of    the T resonance peak frequency, $f_{pk}^T$, of M443 is most easily seen  in the inset to \fig{sepsum}b.   The evidence for coupling of the pinning modes is in the gradual decrease of $f_{pk}^T$ as $p_B$ is decreased over most of its range.  We ascribe the sharp rise of $f_{pk}^T$ near $p_B=0 $ on the plot  to   change in the top-layer density.

 \subsection{Smaller  separations, transition to single pinning mode}
 Samples M417, M433, and M436,  are in the intermediate range of \dt\ in Table 1, and show two  peaks  only for     $p_B$  above a threshold value,  which is denoted  by  $p_B^*$ and is  recorded in Table 1.      The case of imbalance that is small, but sufficient ($\sim 10\%$)  to resolve two resonances when they are present, can give information about the balanced states of the samples.    The $p_B^*$ column in Table 1 indicates that for such small imbalance, samples with $\dt\le 1.8$ (including M433 and M417), keep a single pinning mode, while the samples with larger \dt\  (including M436) show two distinct pinning modes.

%
%
%
%
 \subsection{Single pinning mode: dependence on layer densities} 
 
The smallest \dt\ samples, M465 and M440,  show only one pinning mode in all accessible bilayer states. 
\fig{spx465} shows spectra of M465 with $\tilde{d}=1.56$, for several bilayer and single layer states, at $14$ T.   The spectra are offset upward for clarity, so that the lowest spectrum on the graph is for the largest  $p_{TOT}$. 
The   single resonance   varies in peak frequency, width and intensity as $p_B$ is increased.   The resonance appears sharpest just at balance, and  the peak frequency has a local maximum for  $p_B$ less than $p_T$.


\fig{ptot}a shows $f_{pk}$ and  full width at half maximum,   $\Delta{f}$, vs. $p_{TOT}$ for the resonances of  M465, at two magnetic fields ($14$ and $8$ T). 
For discussion, we mark regions of $p_{TOT}$ on the figure, as well as the balanced state $p_{TOT}$.  For $p_{TOT}$ in region (0), the sample has carriers only in the top layer, so  $p_B=0$, and $p_{TOT}=p_{T}$.   
In region (0) $f_{pk}$  and $\Delta f $ both increase as  $p_{TOT}$  decreases, as was seen in  earlier studies of the pinning modes of  single layer samples \citep{clidensity}.      $f_{pk}$ vs $p_{TOT}$  in region (0) fits $f_{pk}\propto{p}_{TOT}^{-\gamma}$, producing the solid fit lines shown, with $\gamma$  within a typical range of $0.5\pm10\%$, also in agreement with ref. \onlinecite{clidensity}.      

 When $p_{TOT}$ is in  region (1), $f_{pk}$ decreases  with increasing $p_{TOT}$, essentially extending the region (0) power law $f_{pk}\propto{p}_{TOT}^{-\gamma}$. The extrapolation of the power law is shown by dotted lines in \fig{ptot}a.  We do not rule out that owing to charge transfer, the system may remain a single layer with the true bottom-layer density  remaining zero even through region (1), though the increase of the  $\Delta f$  with $p_{TOT}$, seen in the 8 T data,   is  not expected \cite{clidensity} for a single layer system with increasing density. 

     At larger $p_{TOT}$, $f_{pk}$ goes through a minimum; region (2) refers to      the large-$p_B$ side of the minimum, so  has  $f_{pk}$   vs  $p_{TOT}$ with  positive slope.    Within region (2)   $\Delta{f}$ vs $p_{TOT}$ exhibits a local maximum.    Region (3), which is characterized by  $f_{pk}$ vs  $p_{TOT}$ again  decreasing, includes the balanced state.    In region (3), $f_{pk}$ is markedly``pushed up"  above the curve extrapolated from region (0), while $\Delta{f}$ shows a minimum  at the balance point.  
     
  The spectra of the other low-\dt\ samples  M440 and M417 evolve similarly to those of M465, and exhibit the same regions and features  in $f_{pk}$ and $\Delta f$ vs  $p_{TOT}$.   
 
Sample  M433 has slightly larger \dt, which varied from 1.7 to 1.8  between two cool downs, owing to  slightly different as-cooled top-layer densities.   The two \dt\ values give significantly different behavior of $f_{pk}$ and $\Delta f$ vs  $p_{TOT}$, as shown in  \fig{ptot}b.    For $\tilde{d}$ of  1.7, the  intermediate positive slope region (2) of $f_{pk}$ vs $p_{TOT}$ is less prominent than it is for M465. $\Delta{f}$ decreases monotonically on increasing $p_{TOT}$, and does not exhibit any clear feature in region (2) or at balance.   For the slightly higher $\tilde{d}$ of   1.8, a region of $f_{pk}$ vs. $p_{TOT}$ with positive slope is not observed, so that  both $f_{pk}$ and $\Delta{f}$ decrease monotonically on increasing $p_{TOT}$.



\begin{figure}
\clearpage
\centering
\includegraphics[width=2.4in]{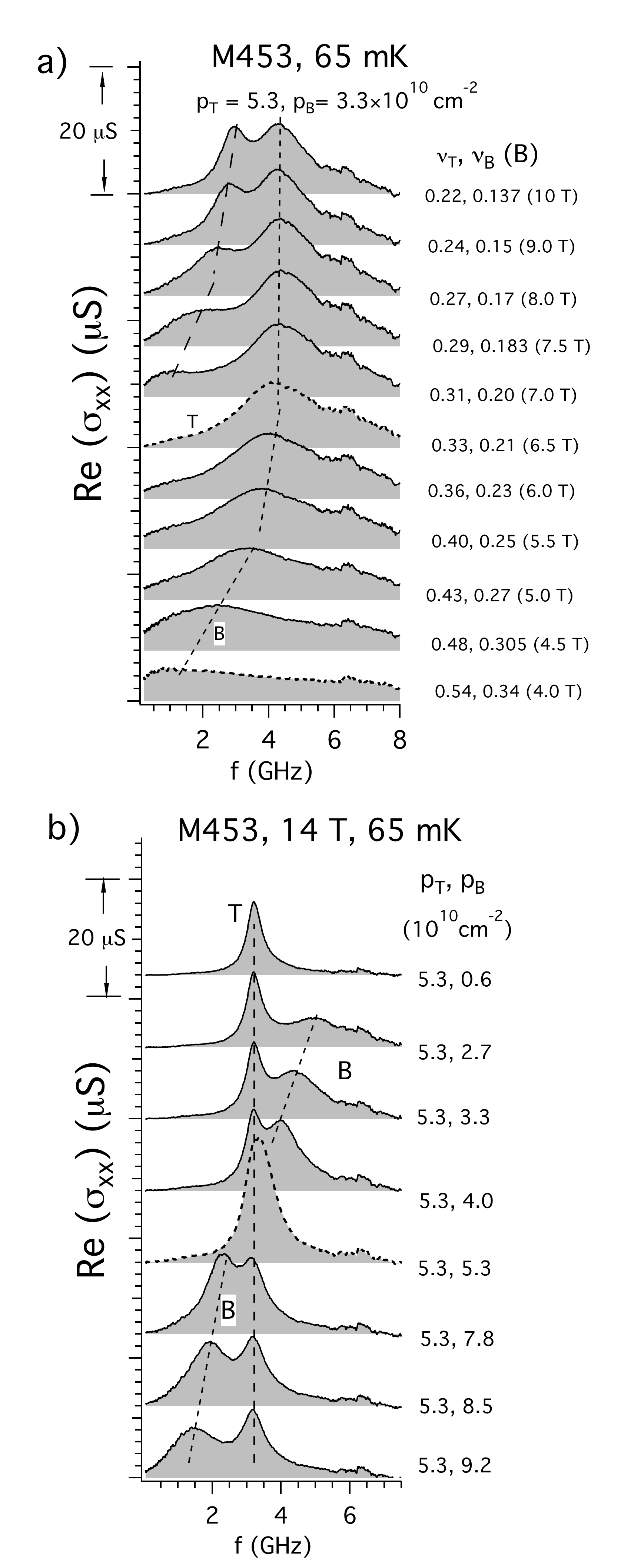}
\caption{(a) Spectra of M453 ($d=2170$ \AA), in a state ($p_T=5.3,  p_B=3.3\times10^{10}cm^{-2}$), at several magnetic fields ($B$), $T\sim65$ mK. The spectra are vertically displaced by $8$ $\mu$S from each other. For each spectrum, filling factors of top and bottom layers, $\nu_T$ and $\nu_B$, are also marked. At $\nu_T=1/3$, or $\nu_B=1/3$, the spectra are plotted as dashed lines. (b) Spectra of M453, in several bilayer states ($p_T$, $p_B$), at $14$ T, $T\sim65$ mK. The traces are offset by $12$ $\mu$S.}\label{figspx453}
\end{figure}

\begin{figure}
\centering
\includegraphics[width=7.5cm]{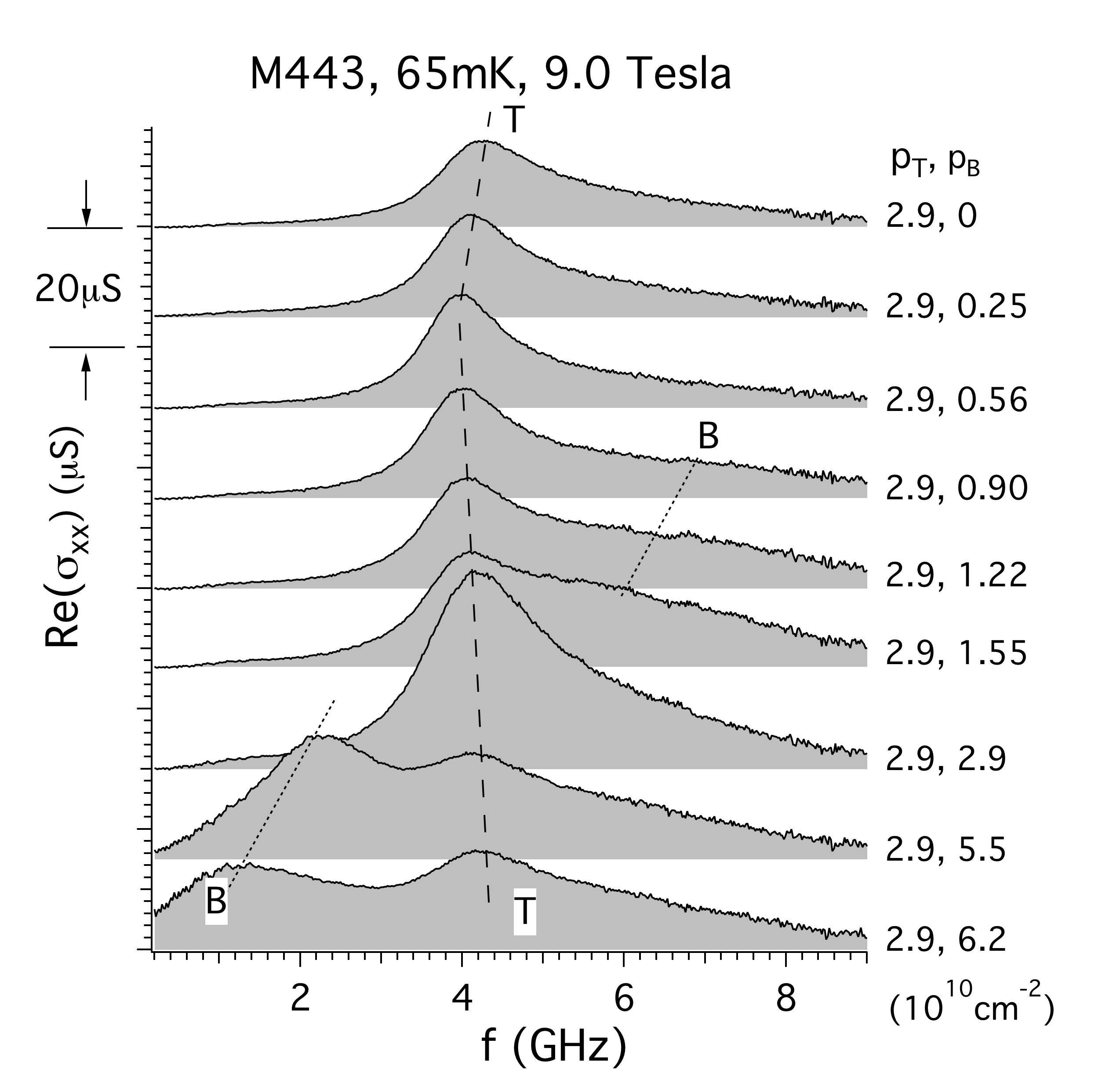}
\caption{Spectra of M443 in several bilayer and single layer states, at $9$ T, $65$ mK.}\label{figspx443}
\end{figure}

\begin{figure}
\centering
\includegraphics[width=7.5cm]{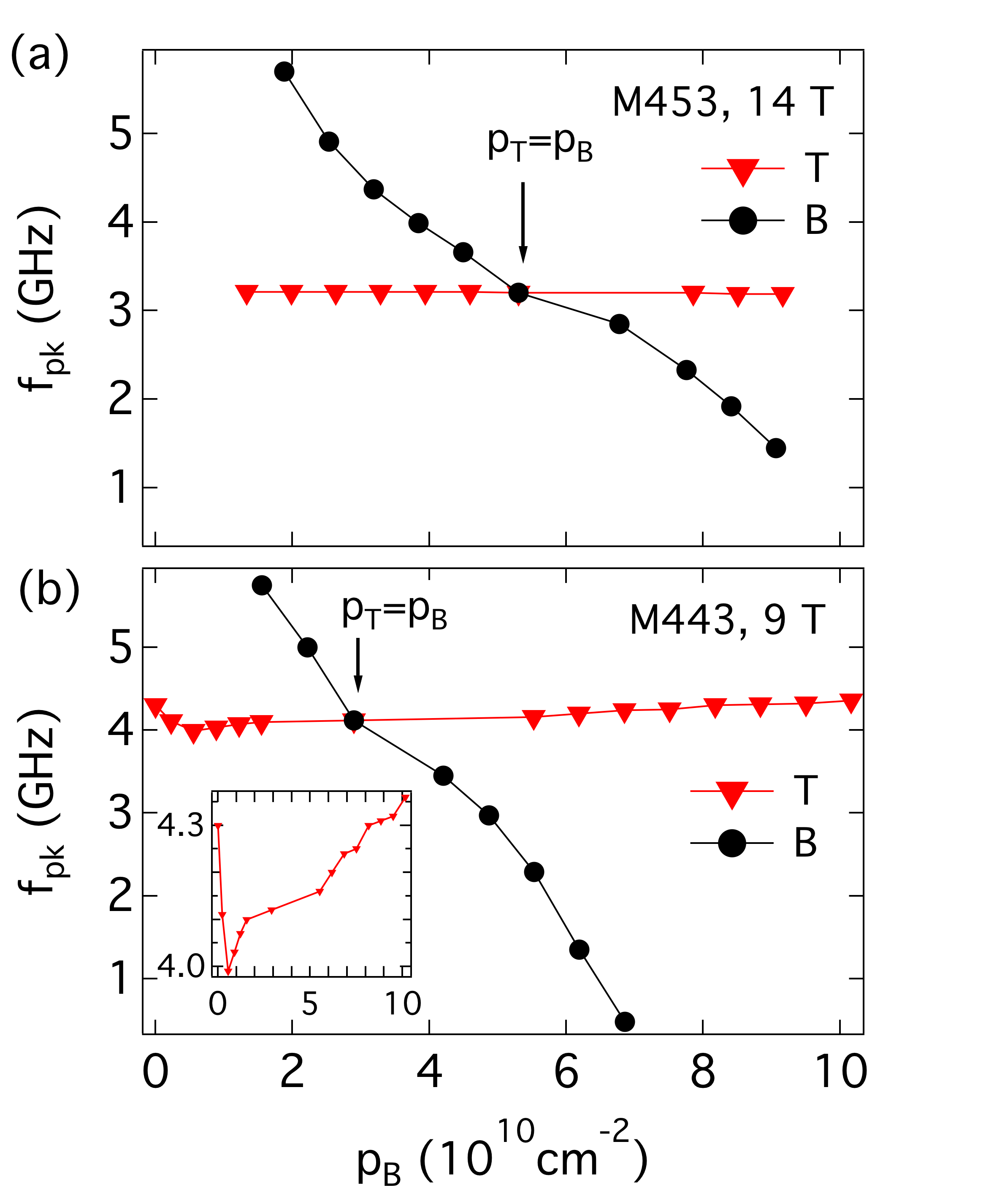}
\caption{(a) $f_{pk}^T$ and $f_{pk}^B$ (peak frequencies of resonances ``T'' and ``B'') vs $p_B$, for M453 ($d=2170$ \AA, $p_T=5.3\times10^{10}{\rm cm}^{-2}$) at $14$ T. The balanced state is marked. (b) $f_{pk}^T$ and $f_{pk}^B$ vs $p_B$, for M443 ($d=650$ \AA, $p_T=2.9\times10^{10}cm^{-2}$) at $9$ T. The inset shows $f_{pk}^T$ vs $p_B$.   
  All data are taken at $T\sim65$ mK.}\label{figsepsum}
\end{figure}

\begin{figure} 
\centering
\includegraphics[width=7.5cm]{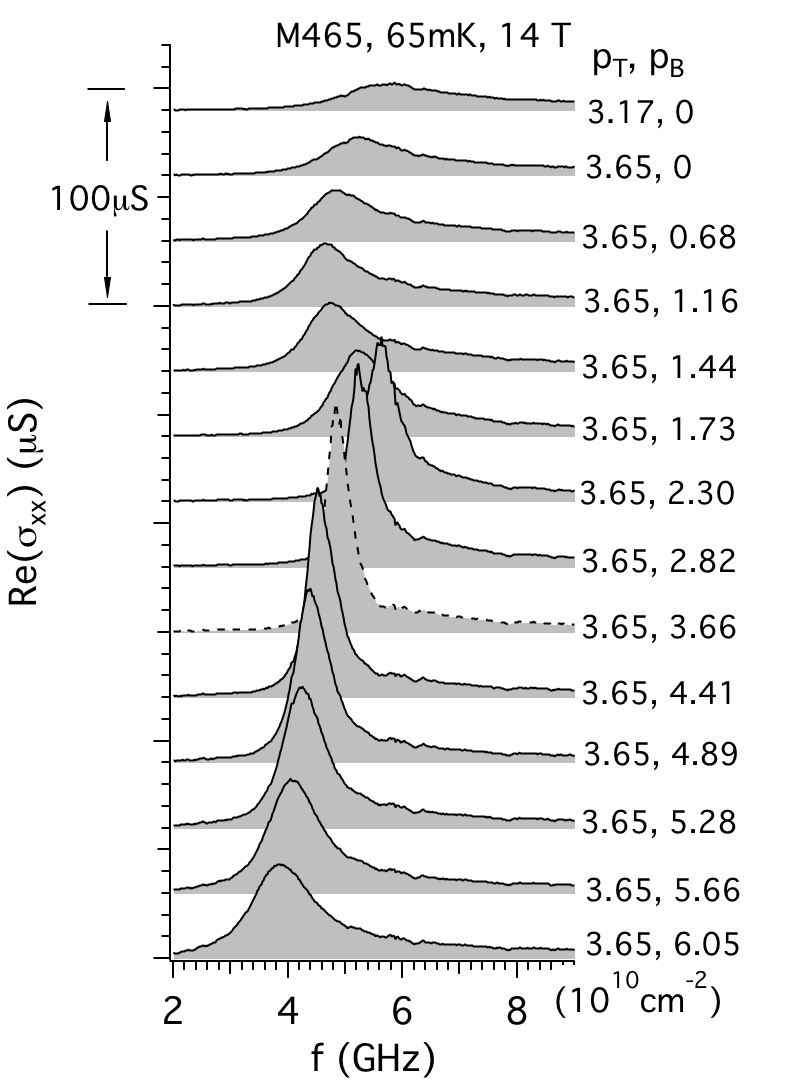}
\caption{Spectra of M465, in several bilayer and single layer states ($p_T$, $p_B$), at $14$ T, $T\sim65$ mK. The spectra are offset by $30\ \mu$S from each other. The spectrum at balance is plotted as a dashed line.}\label{figspx465}
\end{figure}

\begin{figure}
\includegraphics[width=7cm]{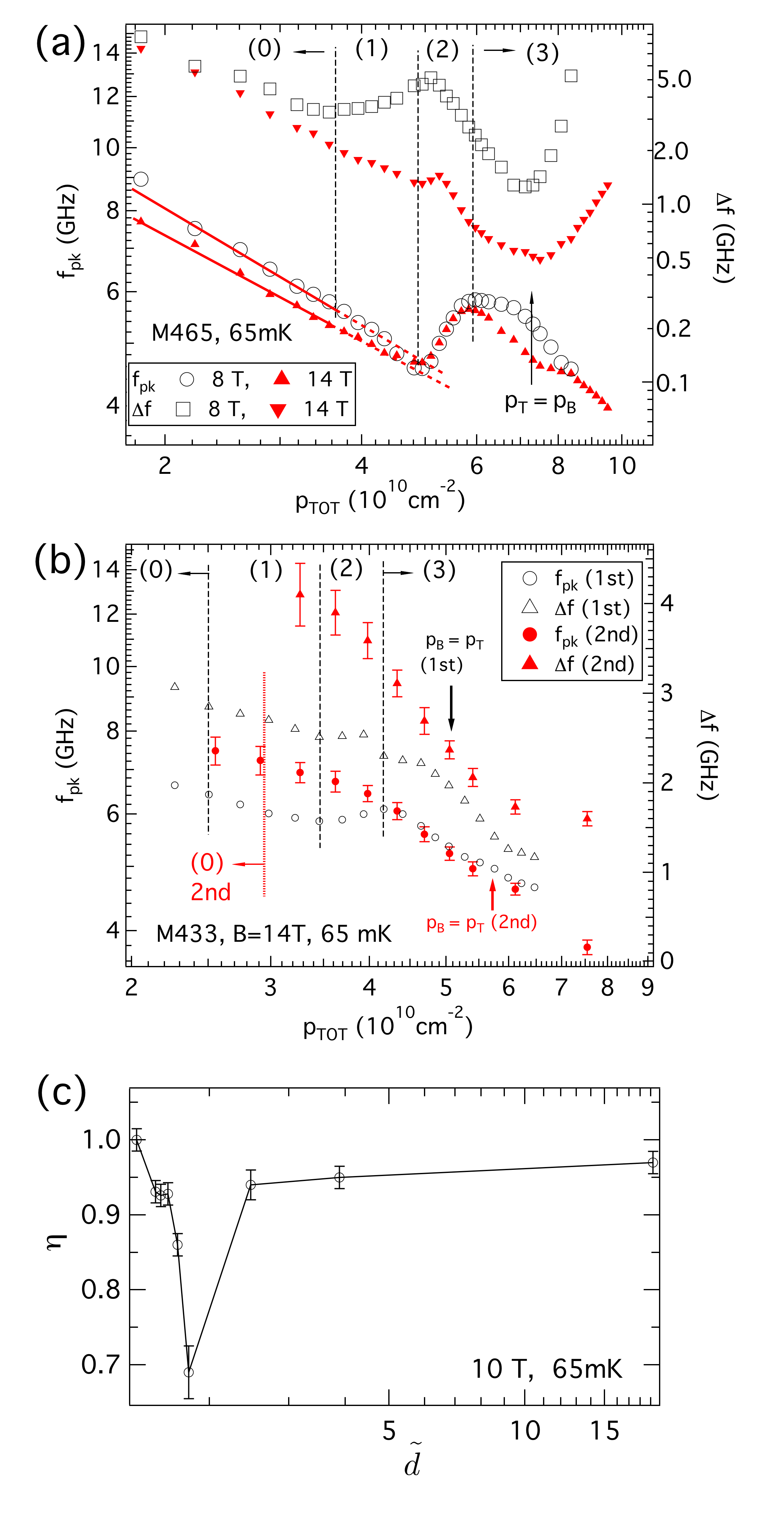}\vspace*{-.6cm}
\caption{(a) $f_{pk}$ and $\Delta{f}$ (full width at half maximum) vs total bilayer density $p_{TOT}$, for M465, at $8$ and $14$ T, $T\sim65$ mK. In region (0), $p_B=0$, $p_T=p_{TOT}$. On the right side of the dashed line, $p_T=3.65\times10^{10}\ {\rm cm}^{-2}$, $p_B=p_{TOT}-p_T$. Three regions (1) - (3), separated by dashed lines, are defined according to the slope of the $f_{pk}$ vs $p_{TOT}$ curve. The balanced state is marked by a vertical arrow.  In regions (0) and (1), $f_{pk}$ vs $p_{TOT}$ fits $f_{pk}\propto{p}_{TOT}^{-\gamma}$, with $\gamma\approx0.5\pm10\%$. (b) $f_{pk}$ and $\Delta{f}$ vs $p_{TOT}$, for two cool downs of M433 ($d=300$ \AA), at $14$ T, $T\sim65$ mK. The two cool downs have different as-cooled top layer densities, leading to different $\tilde{d}$. Black, open symbols, with dashed lines separating regions: $p_T=2.52\times10^{10}\ { \rm cm}^{-2}$, $\tilde{d}=1.7$.  Red closed symbols, with dotted lines separating regions: $p_T=2.85\times10^{10}\  {\rm cm}^{-2}$ and $\tilde{d}=1.8$. (c) Peak frequency ratio, $\eta$,  vs effective separation, $\tilde{d}\equiv{d}/(4\pi{p})^{-1/2}$, where $p$ is the individual layer density in the balanced state. $\eta$ is the ratio of resonances frequency in the balanced state 
$(p,p)$ to that in the state with the bottom layer depleted,  $(p,0)$.   Based on ref  \protect\onlinecite{zhw}. } \label{figptot}. 
\end{figure}

 		\section{Discussion}

The  interpretation of the data is in terms of  two competing effects. The first effect is generic to pinning modes, including those in single layers, and is an increase of \fp\ as the  spacing of the interacting carriers increases.    We will refer to it as the  ``carrier density effect",  since with larger carrier-carrier spacing (smaller density) hence    decreased carrier-carrier interaction, the carriers must associate more closely with the minima in the disorder potential, causing average pinning energy and average restoring force per carrier, and   \fp, to increase.    
   This behavior  has  been established  theoretically \cite{flr,chitra,fogler,fertig}  and  experimentally \cite{clidensity} in single layers. At fixed $B$ within the low $\nu$ carrier solid range, the experiments show   $\fp\propto n_s^{-\gamma}$,     where $n_s$ is the single layer carrier density, and $\gamma\approx 0.5$  at low densities, but $\sim 3/2 $ at higher density. 

The $\gamma\approx 1/2$ power law is seen in \fp\ vs $p_{TOT}$ in regions (0) and (1) in \fig{ptot}, and within region (3) as well.      In region (1), as shown in \fig{ptot}a, $f_{pk}$ vs. $p_{TOT}$ roughly follows the power law  extrapolated from region (0). This  behavior is explainable as due to the carrier-carrier interaction effect only and indicates that the pinning does not change dramatically from the single layer condition of region 0.  This suggests that the carriers do not spread between the two layers, hence the pseudospin does not have an in-plane component.

The second effect is particular to BWCs, and is an enhancement of the pinning in FMBWC relative to other BWC pseudospin orders.   One of us \cite{yongtheory} suggested that when there are positional correlations  between the effective disorders in the top and bottom layers there is an enhancement of \fp\ by a factor of $ 2$ relative to the case of an AFMBWC in the  low-separation limit.  One possible natural source of such correlated disorder would be impurities which enable local interlayer tunneling.   The amount of enhancement relative to the extrapolated carrier-density effect line was considered at balance in ref. 
 \onlinecite{zhw}.    Balanced states $(p,p)$ were compared  to states $(p,0)$ with nearly the same density  in the top layer, but with the bottom layer depleted.   The ratio $f_{pp}/f_{p0}$ of the  pinning mode frequencies   of these states is denoted by $\eta$, which measures the change of pinning on ``adding'' the bottom layer.     That reference presented a curve of    $\eta$  vs \dt, which is reproduced here as \fig{ptot}c.  The curve showed a sharp minimum for $\dt\approx 1.7$.   The theory \cite{ho,zhengfertig} of the FMBWC for disorder-free, balanced bilayers  predicts FMBWC for \dt\ below about $0.4$, 
 considerably below the \dt\ at which we see enhancement of pinning experimentally.   One possible explanation of this discrepancy is that the FMBWC is being stabilized over competing BWC phases by its enhanced pinning.   

Within  region (3), for example in \fig{ptot}a, $f_{pk}$ decreases on increasing $p_{TOT}$, consistent with the carrier density  effect. Overall though, $f_{pk}$ vs. $p_{TOT}$ in region (3) is markedly enhanced above the curve extrapolated from the single layer region (0).   The upward displacement  shows  the pinning is enhanced relative to  the  pinning of single layer Wigner solid seen in region (0) and possibly region (1), after these are corrected for the carrier density effect.
  The enhanced pinning persists, under an imbalance $\vert{p}_T-p_B\vert/(p_T+p_B)$ as large as $20\%$.     The  8 and 14 T data in \fig{ptot} both show the enhancement, with the 8 T enhancement slightly larger around the  balanced state.  This remains  consistent with both fields being sufficient to produce well developed BWC  solids; the differences between the 8 and 14 T curves  may involve the dependence of effective disorder---which incorporates the in-plane extent of a carrier wave function---on the magnetic length\cite{yewc}. 

The  ``pushed up" region (3) is observed only for states on the low \dt\ side of the minimum in \fig{ptot}c.   We interpret    the \fp\ enhancement in this region as due to   FMBWC formation, as was presented for the balanced case in 
Ref.  \onlinecite{zhw}, but also extending into imbalanced states. This is possible since
even within an FMBWC,  the carrier wave function may not spread equally in the two layers.   In  pseudospin language, this would also mean that at least some pseudospins have an in-plane component, and  the orientation of the pseudospins   changes with density imbalance.    Within region (3),  $f_{pk}$ shows no sharp feature at balance, but $\Delta{f}$ does exhibit a minimum  just at balance. This suggests an imbalance-induced effect, which increases damping of the resonance without affecting pinning. Based on the classification of BWC at balance, this damping is present only when the BWC at balance is an easy-plane ferromagnet.  The damping may result from excess carriers of the majority layer, which we speculate may act as defects, or could themselves form a condensed phase\cite{kunimbal}.

%
 


Region (2) can be interpreted as a transition between   between region (1) with pinning of individual layers (smaller $f_{pk}$) and region (3) with enhanced pinning (larger $f_{pk}$). The local $\Delta{f}$ maximum, which is observed in the middle of region (2), could then be interpreted as due to a transition between different BWC phases. At the transition region, multiple BWC phases can in principle coexist. The broadening effect possibly originates from dissipative excitations that are associated with the phase boundaries.  This picture is strengthened by the data on  M433 in 
\fig{ptot}b which show  no  local $\Delta{f}$ maximum.
For  M433, which has  $\tilde{d}=1.8$,      \fp\  enhancement  in  region (3) is     much less than that in  M465, which has lower \dt.    Instead, $\Delta{f}$ for M433 decreases on increasing $p_{TOT}$, which can be interpreted as due to the carrier density effect.

 An alternate explanation of the enhancement of \fp\ around balance in region (3) in \fig{ptot}a  (and the minimum in $\eta$ vs \dt\ in \fig{ptot}c) is that there is an increase in \fp\ due to softening of the BWC.  Such softening might occur 
 around  a phase transition from one type of BWC to another, since near a transition many carrier arrangements have similar energies.   The $\Delta f$ data in   
 \fig{ptot}a  are difficult to reconcile to  this picture, in which  region (3) rather than region (2) would be identified with a transition, since the maximum in $\Delta f$ occurs in region (2), and a minimum in $\Delta f$ occurs in region (3).  If softening of the solid occurs at a phase transition, one would naturally expect more damping at the transition,  contrary to the data.  
 
The behavior of the samples at small  imbalance  (small but sufficient to resolve the two pinning modes of the independent layer case)  is a strong indication that a balanced sample would  go though a phase transition as \dt\ goes below 1.8 (the minimum in \fig{ptot}c).  For $\dt\le 1.8$  there is  only one resonance at small imbalance,  while for  $\dt> 1.8$ the resonance splits.

In summary,  by studying the peak frequency in the imbalanced states, we found that the enhanced pinning exists not only at balance, but also over a considerable range of imbalanced states (around the balance)  for low-\dt\  bilayers. 
In addition, the small-separation samples that show the enhanced pinning  do not exhibit a splitting of the pinning mode when the layer densities are subject to small imbalance.   The interpretation is in  terms of a pseudospin FMBWC, 
 present at balance but  persisting  even for considerable imbalance.

This work was supported by DOE Grant Nos. DE-FG21-98-ER45683 and  DE-FG02-00-ER45841 at Princeton, and  DE-FG02-05-ER46212 at NHMFL. NHMFL is supported by NSF Cooperative Agreement No. DMR-0084173, the State of Florida and the DOE.
 

\end{document}